# Spreading of families in cyclic predator-prey models

Mária Ravasz
*Babeş-Bolyai University, Department of Physics, RO-400084 Cluj, Romania*

György Szabó and Attila Szolnoki
*Research Institute for Technical Physics and Materials Science P.O. Box 49, H-1525 Budapest, Hungary*
(Dated: April 29, 2004)

We study the spreading of families in two-dimensional multispecies predator-prey systems, in which species cyclically dominate each other. In each time step randomly chosen individuals invade one of the nearest sites of the square lattice eliminating their prey. Initially all individuals get a family-name which will be carried on by their descendants. Monte Carlo simulations show that the systems with several species ($N = 3, 4, 5$) are asymptotically approaching the behavior of the voter model, i.e., the survival probability of families, the mean-size of families and the mean-square distance of descendants from their ancestor exhibit the same scaling behavior. The scaling behavior of the survival probability of families has a logarithmic correction. In case of the voter model this correction depends on the number of species, while cyclic predator-prey models behave like the voter model with infinite species. It is found that changing the rates of invasions does not change this asymptotic behavior. As an application a three-species system with a fourth species intruder is also discussed.

The original Lotka-Volterra model [1, 2] describes the struggle for existence among interacting (homogeneous) populations such as predator-prey systems. The cyclic variant of the model is attracting increasing theoretical interest. The reason is that dynamically equivalent absorbing states and spatial structure eventuate a rich variety of interesting phenomena in the dynamics of the model [3–7]. Spreading phenomena are frequently studied in models with absorbing states. The best known examples are the contact process (for review [8, 9]), voter model [10], but exploration of models with more absorbing states is far from being complete.

In the present model each site of a square lattice is occupied by one of the $N$ species dominating cyclically each other. They mimic a cyclic food chain, that is each species has one predator and one prey (e.g., A eats B eats C eats A for $N = 3$). The following sequential dynamics is implemented: *i)* we randomly choose a site and one of its nearest neighbors; *ii)* if they are different species the predator eats the prey, by that we mean the predator duplicates itself and the prey is eliminated ($A + B \to 2A$). Nothing happens if they are neutral species which do not interact (this can be only if $N > 3$). The simulations are performed on a square lattice with $L \times L$ sites under periodic boundary conditions. In a time unit, called a Monte-Carlo step (MCS), each site of the lattice is chosen once on the average. In such a spatial system the variation of species distribution can occur only along boundaries separating a predator-prey pair. Consequently, both the single species (homogeneous) states and the mixtures of neutral species are considered as absorbing states (where the system stays forever once it is reached). Starting from a random initial state this system evolves towards a self-organizing pattern of species distribution that is maintained by the cyclic invasions [5, 6] if the system is sufficiently large for $3 \le N \le 14$. Frachebourg *et al.* [3] have shown that fixation occurs on arbitrary large square lattice if $N > 14$. Henceforth our analysis will be restricted to the cases $N = 3, 4$, and 5.

In this Brief Report we study the two-dimensional spreading and extinction of families (or colonies) in multispecies predator-prey systems. Spreading of epidemics are usually studied by starting from a single active site initial condition [8]. In such a situation all the active sites have the same common ancestor or origin, so they are members of a single family. The weakness of this procedure is that many independent realizations ($10^6$-$10^7$) have to be made for good statistics. Now our simulations are started with a random initial state and after a suitable thermalization time (needed to achieve the stationary state in $t = 0$) all the individuals on the sites of the lattice get different "family names" which will be inherited by the descendants (for example if $A$ with name 1 eats $B$ from family 2 the new-born $A$ inherits his parent's family name 1). Subsequently, sites with identical names have a common ancestor and form a family. Thus we can follow simultaneously the history of $L \times L$ families and evidently this way improves the statistics.

The growth of families is customarily described by the average survival probability of the families $P(t)$. A family survives if at least one individual on the lattice carries that name. As a consequence, $P(t)$ is the number of surviving families normalized by $L \times L$. Further measured quantities are the mean population size of the families $n(t)$ and the mean-square distance of descendants from their original ancestor $R^2(t)$ characterizing the average spreading of the surviving families [8].

In some cases the scaling of these quantities was found to be $P(t) \propto t^{-\delta}$, $n(t) \propto t^\eta$, and $R^2(t) \propto t^z$ [8, 9]. In the present model, however, the mean population size remains unchanged [$n(t) = 1$ corresponding to $\eta = 0$] because the total number of surviving individuals is conserved for each elementary invasion. The classical voter models [10–12] possess the same feature. This similarity has inspired us to compare the above predator-prey systems with the voter models from the viewpoint of family growth.

In the voter model the individuals distributed on a lattice represent one of the $N$ opinions that are modified by random sequential updates. More precisely, if two randomly chosen, nearest-neighbor individuals have different opinions (say $A$ and $B$) then the invasions in both directions, $A + B \to 2A$ or $A + B \to 2B$, take place with the same probability. In case of two opinions ($N = 2$) we get back the classical voter model [10–12]. First our numerical investigations were focused on a particular case when all the individuals have different opinions ($N = L^2$) at $t = 0$. Notice that the family growth in this model agrees with those studied by Dickman and Tretyakov [13] considering the spreading from a single 1 in the sea of 0s. Due to the enhanced accuracy of the applied method we can demonstrate the occurence of the logarithmic correction to scaling of the survival probability $P(t)$.

Several studies confirm that scaling of different quantities have logarithmic corrections in the two-dimensional voter model [10, 14, 15]. The most common example is the density of interfaces: $\rho \propto 1/\ln t$ [15]. The assumption of logarithmic correction is also beneficent for the scaling of the survival probability of families. Averaging over ten simulations on a $2000 \times 2000$ lattice (it means that quantities were averaged on $4 \cdot 10^7$ families) up to a maximum time of $10^5$ MCS have resulted in that the survival probability of the families can be described by the expression $P(t) \simeq (a + b \ln t)/t$ (Fig. 1b) more exactly, than by a simple power law used in the latest studies (Fig. 1a). The numerical fitting gives $a = 0.487(2)$ and $b = 0.302(1)$.

These simulations were repeated with choosing randomly one of the $N = 2$, or 3 opinion (species) labels for the $L^2$ families at $t = 0$. In contrary with the case $N = L^2$ here we have more families with the same opinion, but in the evolutionary process only those family (and species) invasions are permitted where the nearest neighbors belong to different species. This constraint yields slower extinctions. In the asymptotic region ($t \gtrsim 100$ MCS) the results of numerical simulations for $P(t)$ are consistent with a similar behavior as above. In this case, however, the fitted values of the parameters are changed and as shown on Figure 1b the slope depends on the number of species ($N$). This dependence follows from the fact that invasions are permitted only between different species, so an individual can react only with the $(N-1)/N$ fraction of the whole population, this means the survival probability of families declines more slowly if $N$ is small. Fig. 1c proves that the slopes can be approximated with the following relation $b(N) = b(\infty) * N/(N-1)$, where $b(\infty) = 0.302(1)$ is the slope calculated above (when all individuals had different opinions in $t = 0$, which corresponds to $N \to \infty$).

Comparing the numerical results we obtained for the cyclic predator-prey systems and the results for the voter model (presented above) we have noticed that in cyclic predator-prey systems after a transient time the quantities $P(t)$ approach asymptotically to a similar function given above. The length of the transient period increases with the number of species $N$, so the asymptotic behavior could be detected only in case of systems with $N = 3, 4,$ and $5$. The asymptotic slopes ($b$) seem to be the same as for the voter model with infinite species (Figure 2). In order to avoid the undisered size effect [in the measurement of $R^2(t)$] for $N = 3, 4$ the simulations were performed on a $2400 \times 2400$ lattice (20 respectively 10 realizations) up to $10^5$ MCS, for $N = 5$ the size of lattice was $L = 3000$ (10 realizations) up to time $10^5$. For systems with $N > 5$ the preliminary simulations suggest the same behavior but for conclusive results simulations up to more timesteps should be made. It is interesting that during the transient time the quantity $tP(t)$ goes through a maximum which is increasing with $N$. This observation is related to the fact that the probability for an individual to be surrounded by neutral partners grows with $N$, so the family extinction is blocked. In the asymptotic period, however, the slope does not depend on the number of species and these systems behave like the voter model with infinite number of species. In this case the individuals typically meet their predator or prey at the boundary of a domain of family and the invasion is not blocked anymore as well as in the voter model for $N \to \infty$.

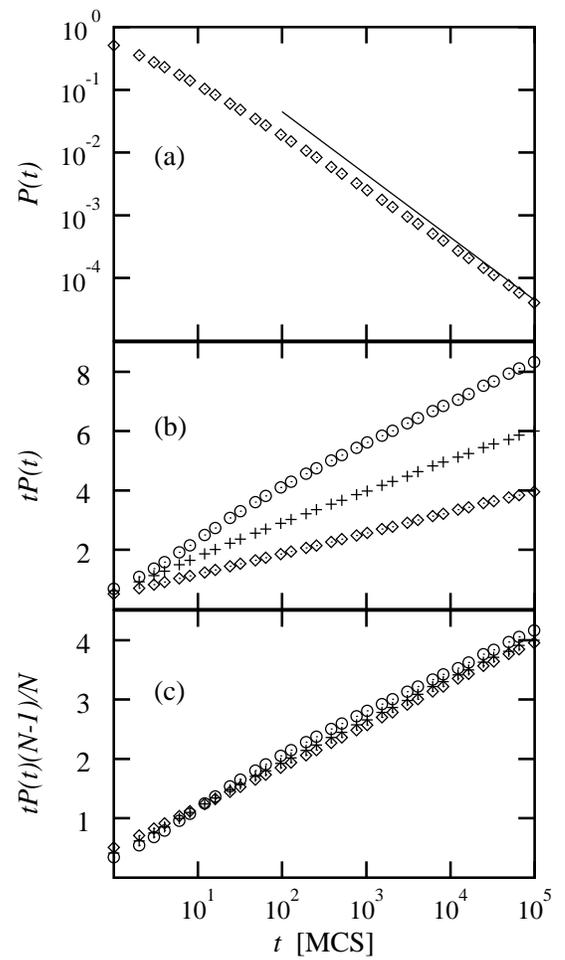

FIG. 1: MC results for the voter model. a) Log-log plot of the survival probability as a function of time. The asymptotic $t^{-1}$ function is denoted by a solid line. b) Log-lin plot of the survival probability of families multiplied by time as a function of time. c) The same log-lin plot, but $tP(t)$ is now multiplied by $(N-1)/N$. The open diamonds indicate the voter model with infinite species, when all individuals represent different families and species at $t = 0$ ($N \to \infty$). Results for the voter model with $N = 2$ and $3$ species are denoted by open circles and plus signs respectively.

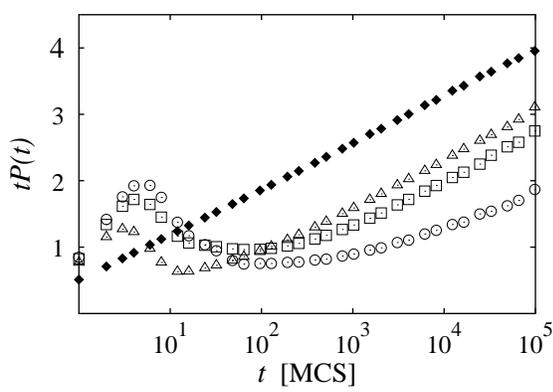

FIG. 2: Log-lin plot of the survival probability of families multiplied by time as a function of time. The closed diamonds indicate the MC results for the voter model if the $L^2$ individuals represent different families and species at $t = 0$ ($N \to \infty$). For the cyclic predator-prey systems the results are denoted by open triangles ($N = 3$), open squares (4), and open circles (5).

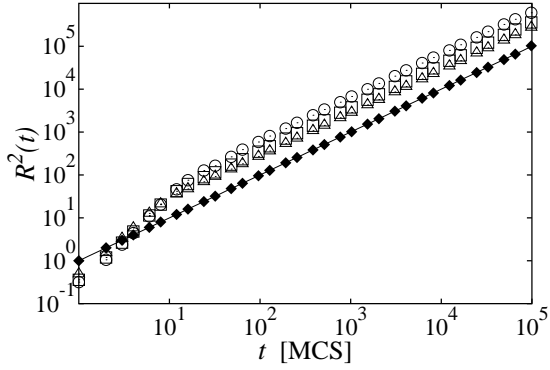

FIG. 3: MC results for the mean-square spread of individuals about their ancestors in function of time for the voter model with infinite species and cyclic systems with $N = 3$, 4, and 5 species. Symbols as in Fig. 2.

The exponent of $R^2(t)$ was also found to be the same as in the voter model independently of the value of $N$. Figure 3 demonstrates clearly that $R^2$ grows linearly with time in the asymptotic region. The numerically fitted exponents agree with the theoretical prediction ($z = 1$) within the statistical error that is less than one percent. In the voter model this behavior is explained by its relation to the random walk [10, 14].

This voter-like asymptotic behavior shows that after particles spread out far enough in time and space about their origin in the fight of two individuals the winner turns out to be chosen randomly (far from his ancestor the individual already has the same chance to meet his predator as his prey), like in the voter model, where the winner could be each of the individuals with equal probabilities. This assumes that correlations decrease rapidly. Indeed measuring the correlation function in case of $N = 3$ (sometimes called rock-scissor-paper system, shortly RSP) on a $2000 \times 2000$ lattice we have observed it is decreasing exponentially and the correlation length being $\xi = 2.54(5)$.

Another frequently measured quantity is the survival probability of the individuals, or in the language of the voter model the fraction of persistent voters, who did not change their opinion until time $t$ (in our model the number of still living ancestors). Measuring this quantity on the same system supports that it is also decreasing exponentially with a relaxation time of 1.87(5) MCS.

In these cyclic predator-prey models all species play equal roles and represent dynamically equivalent absorbing states. One could ask how a minor braking of symmetry influences the behavior of the system. We studied an RSP system in which one of the rate of invasions was changed, the probability that this invasion happens being only 0.8 while the probability of other invasions remained the same (1). Simulations show that the equilibrium concentrations of species change, meanwhile the behavior of the system (scaling of $P(t)$, $n(t)$, $R^2(t)$) remains the same as presented above.

As another interesting application we studied a three-species model (RSP) with an intruder of type $D$ substituted for a randomly chosen individual after the stationary state of the RSP system is achieved. The intruder represents a fourth race having one prey and one predator among the other three species. Considering symmetries one can easily recognize that there are only two different structures of the foodweb, when the prey of the intruder is the predator of the neutral partner of the intruder ($A \to B \to C \to A$, $A \to D$ and $D \to B$) and the second case when the intruder has the same predator and prey as his neutral partner (the race with whom it does not interact), ($B \to D$ and $D \to A$).

In the first case family of the intruder rapidly dies out, the survival probability $P(t)$ as well as the mean size of colonies $n(t)$ decreasing exponentially. $5 \cdot 10^5$ realizations on a $200 \times 200$ lattice up to time 200 MCS starting with a single intruder resulted in the relaxation time of $P(t)$ being 35(1) MCS and for $n(t)$ 36(1) MCS. For the scaling of $R^2(t)$ we get $z = 1.01(2)$ showing that motion of the few survivors can well be described by simple random-walks. In this case we could say the RSP system forms a "defensive alliance" against the intruder [16].

In the second case the intruder has the same predator and prey as his neutral partner, so his behavior is also the same, ($P(t) \propto (a + b\ln(t))/t$, $R^2(t) \propto t$), and only the mean size of the intruder family changes, asymptotically approaching to $n(t) \approx 3$ (results from 85400 simulations on a $520 \times 520$ lattice up to time $10^3$). The fact that the mean-size of intruder families is higher than 1 (mean-size of families in simple RSP systems) suggests the succesors of the intruder (who just randomly took the place of an individual while eliminating it), enjoy a benefit during the evolution, like the system would prefer this kind of migrations.

In summary, our MC results point out that from the aspect of spreading phenomena cyclic predator prey systems show a voter-like asymptotic behavior. Following the survival probability of families, the mean-size of the families and the mean-square distance of spreading from their ancestors we observe that they asymptotically approach the same scaling as in the voter model ($P(t) \simeq (a + b\ln t)/t$, $n(t) = 1$, $R^2(t) \propto t$). The survival probability has a logarithmic correction, but while for the voter model the parameters of this correction depend on the number of species, for the cyclic predator-prey systems this dependence is missing, and they asymptotically aproach the behavior of the voter model with infinite species. The most common feature of these cyclic models and the voter model is the absence of surface tension and the dynamics driven only by interfacial noise between dynamically equivalent absorbing states (invasions take place only at the borders separating homogeneous domains of species). For $N = 3$ the cyclic symmetry does not seem to be so important for this behavior as we saw, slightly changing the rates of invasions, only the concentrations of species changed, while the spreading of colonies remained the same. For a long time the voter model was considered to be a marginal system because of the exceptional character of its analytic properties, but studies have shown that in fact it represents a broad class of models which show the same type of coarsening phenomena [14]. Now studying spreading processes we find again that a whole class of models (whose evolution is governed by quite different local rules) show the same behavior as the voter model, and common features on large scales are again the same: absence of surface-tension and interfacial dynamics.

The last application about colonies of intruders (specially the second case) brings to light an interesting feature of the system. Because of special initial conditions (the intruder is dropped into the RSP system, taking the place of another individual) the intruder has a bigger chance to find his preys than his neutral partners which are grouped together (most of them being surrounded by neutral individuals), so the mean-size of intruder colonies turns out to be higher than the mean-size of families in normal systems, showing that in this cyclic system this kind of migrations guarantee evolutionary a point of vantage. Although ecological systems are not as simple as our models, it is interesting to notice that indeed different kind of migrations of animals occur very often in nature.

### Acknowledgments

This work was supported by the Hungarian National Research Fund under Grant Nos. T-33098, F-30499, and Bolyai Grant No. BO/0067/00.